# Enabling Privacy-Preserving, Compute- and Data-Intensive Computing using Heterogeneous Trusted Execution Environment

Jianping Zhu[1], Rui Hou[1,*], XiaoFeng Wang[2], Wenhao Wang[1], Jiangfeng Cao[1], Lutan Zhao[1],
Fengkai Yuan[1], Peinan Li[1], Zhongpu Wang[3], Boyan Zhao[3], Lixin Zhang[3], Dan Meng[1]

[1]*State Key Laboratory of Information Security, Institute of Information Engineering, CAS
and University of Chinese Academy of Sciences.*
[2]*Indiana University Bloomington.*
[3] *Institute of Computing Technology, CAS.*

## Abstract

There is an urgent demand for privacy-preserving techniques capable of supporting compute- and data-intensive (CDI) computing in the era of big data. Answering to this urgent call are secure computing techniques, which have been studied for decades. Compare with traditional software-only approaches such as homomorphic encryption and secure multi-party computing, emerging as a more practical solution is the new generation of hardware supports for Trusted Execution Environments (TEEs). However, none of those existing TEEs can truly support CDI computing tasks, as CDI requires high-throughput accelerators like GPU and TPU but TEEs do not offer security protection of such accelerators. This paper present HETEE (Heterogeneous TEE), the first design of TEE capable of strongly protecting heterogeneous computing with unsecure accelerators. HETEE is uniquely constructed to work with today's servers, and does not require any changes for existing commercial CPUs or accelerators. The key idea of our design runs *security controller* as a stand-alone computing system to dynamically adjust the boundary of between secure and insecure worlds through the PCIe switches, rendering the control of an accelerator to the host OS when it is not needed for secure computing, and shifting it back when it is. The *controller* is the only trust unit in the system and it runs the custom OS and accelerator runtimes, together with the encryption, authentication and remote attestation components. The host server and other computing systems communicate with *controller* through an in-memory task queue that accommodates the computing tasks offloaded to HETEE, in the form of encrypted and signed code and data. Also, HETEE offers a generic and efficient programming model to the host CPU. We have implemented the HETEE design on a hardware prototype system, and evaluated it with large-scale Neural Networks inference and training tasks. Our evaluations show that HETEE can easily support such secure computing tasks and only incurs a 12.34% throughput overhead for inference and 9.87% overhead for training on average.

---

*Corresponding author: Rui Hou (hourui@iie.ac.cn)

## 1 Introduction

The explosive growth of the data being collected and analyzed has fueled the rapid advance in data-driven technologies and applications, which have also brought data privacy to the spotlight as never before. A large spectrum of data-centric innovations today, ranging from personalized healthcare, mobile finance to social networking, are under persistent threats of data breaches, such as Facebook data exposure [10,11], and the growing pressure for compliance with emerging privacy laws and regulations, like GDPR (general data protection regulation) and the CCPA (California Consumer Privacy Act). As a result, there is an urgent demand for privacy-preserving techniques capable of supporting compute- and data-intensive (CDI) computing, such as training deep neural networks (DNNs) over an enormous amount of data.

**Protecting CDI computing**. Answering to this urgent call are *secure computing* techniques, which have been studied for decades. Traditional software-only approaches such as homomorphic encryption and secure multi-party computing are considered to be less effective in protecting complicated computing (such as DNN analysis) over big data, due to their significant computation or communication overheads. Emerging as a more practical solution is the new generation of hardware supports for *Trusted Execution Environments* (TEEs) such as Intel Software Guard Extensions (SGX) [25], AMD Secure Encrypted Virtualization (SEV) [24] and ARM TrustZone [3]. These TEEs are characterized by their separation of a secure world, called *enclave* in SGX, from the insecure one, so protected data can be processed by trusted code in an enclave, even in the presence of a compromised operating system (OS) and corrupted system administrators.

However, none of those TEEs can truly support CDI computing tasks, due to their exclusion of high-throughput accelerators such as graph-processing unit (GPU), tensor-processing unit (TPU), field-programmable gate array (FPGA, for building custom accelerators) etc. More fundamentally, today's TEEs are *not* designed to protect big-data analytics, since they fail to support the *heterogeneous computing* model that becomes the mainstream architecture for CDI computing [2,9,18,19,22,38,42,43]. Under a heterogeneous architecture, a CDI computing task is jointly processed by different types of computing units, such as CPU, GPUs, cryptographic co-processors and others. For example, a machine learning task today is typically performed by a CPU, which acts as a control unit, and a set of GPUs or TPUs, which serve as computing units. This collaborative computing model is shown to significantly outperform a homogeneous one [19,38]. In the meantime, such collaborations need to be under a TEE's protection during secure computing, which has not been considered in the current designs[1]. A related issue is these TEEs' Trusted Computing Base (TCB), whose software and hardware stacks range from nothing but CPU (Intel SGX), thereby forcing a computing task to use OS resources and expose itself to side-channel threats, to a generic OS running on multiple trusted units (e.g., MMU, DMA and other controllers supporting ARM TrustZone, in addition to CPU) with the level of complexity that likely introduces vulnerabilities. A TEE for CDI computing is expected to strike a balance between attack surface reduction and TCB simplification, by only managing the resource for computation.

Recently an attempt has been made to extend TEE protection for CDI computing [53], through adding trusted computing components to GPU chips. The approach, however, still does not support the heterogeneous model. Actually, in order to work with CPU, a secure channel needs to be established between a CPU enclave and the modified GPU, which is exposed to side-channel analysis by untrusted OS and incurs additional performance overheads. Also, this design requires changes to the architectures of GPU chips, such an expensive process may also introduce new challenges in establishing trust across different vendors' computing units, when they are used together to support a heterogeneous computing task.

**Heterogeneous TEE**. We believe that a TEE designed for CDI tasks should offer a strong support for heterogeneous computing, enabling collaborative computing units to be protected under a single enclave and conveniently assigned across secure/insecure worlds and different enclaves. Further this computation-oriented TEE should only include a small yet computing-ready TCB (e.g., a single *security controller* with a software stack customized for supporting CDI tasks), to reduce its complexity and also minimize the side-channel attack surface exposed by resource sharing with the untrusted OS. For ease of deployment, using existing computing units without chip-level changes is highly desirable. In this paper, we present the first design that meets all these descriptions. Our approach, called *HETEE* (*Heterogeneous TEE*), is uniquely constructed to work with today's servers. It includes a *security controller* running on a board connected to the standard PCI Express (or PCIe) bus, acting as a gatekeeper for a secure world dedicated to CDI computing. The unit can be built using an FPGA board (the FPGA chip embodies ARM processor cores and programmable logics) or a custom board with a commodity processor (e.g., X86 processor), and is the *only* unit that needs to be trusted. It runs a thin software stack to operate accelerator drivers and computing platforms like CUDA, together with secure computing functions like encryption/decryption, authentication and attestation (to the data holder). All the computing units, like GPUs, AI accelerators, etc., are *shared* between the OS and the *security controller*. *A unique design of HETEE is its use of a PCIe switch (such as PCIe ExpressFabric series chip produced by Broadcom) to dynamically move computing units between the secure world under its control and the insecure world managed by the OS*. The switch also allows the *security controller* to partition the secure world into different *enclaves* with cross-enclave communication under its full mediation.

A critical issue here is how to trust the computing units also used in the insecure world. Our solution leverages the unique features of today's GPU and accelerator designs: most of them are tight, fully controlled by firmware and carrying a thin software stack; so they can be "reset" to a trusted status by simply cleaning up all their internal storage. Particularly, depending on the type and vendor of a computing unit, HETEE either utilizes its "hot reset" function to restore it to the clean status or runs a program to wipe out all its storage content. We show that this simple approach works effectively on mainstream units like NVIDIA GPU, Tesla M40 and GTX TITAIN X. In this way, HETEE avoid any chip-level changes and can work with all existing computing units, and therefore can be easily deployed to servers to enable large-scale computing.

We partially implemented our design on the prototype system which is a X86 host system connected with Xilinx Zynq FPGA (acting as the *security controller*) and NVIDIA GPUs over PCIe fabric, and evaluated it with large-scale Neural Networks inference and training tasks. Our study shows that compared with unprotected computation on GPUs, HETEE can easily handle the task involving big data and only incurs a 12.34% throughput overhead for inference and 9.87% overhead for training on average. We also discuss how to protect the design from physical attacks, using existing tamper-resistant techniques.

**Contributions**. The contributions of the paper are summarized as follows:

- *First heterogeneous TEE design*. We present the first design of TEE capable of protecting heterogeneous computing, which enables different computing units (CPU, GPU, accelerators, etc.) to efficiently work together, without exposing a large attack surface to untrusted OS. Further, by using a PCIe switch and state reset, our design can conveniently and safely move computing units in and out of the secure world, effectively compartmentalizing a secure computing task with limited additional resources required from current computing systems.
- *Ease of deployment and immediate impacts*. Our solution does not require changes to accelerators' architectures, and can therefore be relatively easy to produce and deploy, providing immediate protection to today's CDI computing.
- *Implementation and evaluation*. We partially implemented our design and evaluated it on real-world CDI tasks, which demonstrates the effectiveness of our new solution.

---

[1] Note that though theoretically the secure world of TrustZone can be extended to peripherals, including a GPU unit, in practice, ARM tends to move GPUs outside the protection [3], not to mention proving the trustworthiness of a computing task running in the unit through attestation.

**Roadmap**. The rest of the paper is organized as follows: Section 2 is the background; Section 3 introduces the HETEE design and make security analysis; Section 4 presents the performance and cost evaluations on the hardware prototype system; Section 5 discusses the limitations of the current approach and potential future research; Section 6 is the related work and Section 7 concludes the paper.

## 2 Background

### 2.1 Heterogeneous System

Heterogeneous system has been considered to be a promising solution to CDI computing, due to its potentials to achieve high performance at a low energy cost. A prominent example is the wide adoption of the CPU+GPU combination for the computing tasks ranging from machine learning model training and inference to the support for automatic driving. Also, Google uses its in-house TPU chips for AI acceleration, and Microsoft has deployed thousands of FPGAs to speed up its Bing search service. Central to such a heterogeneous system are its computing units, such as accelerators, which can be characterized by the following features:

**Producer-consumer based programming model**. Behind all kinds of accelerators is the same programming model in which the host program running on CPU acts as a consumer and the accelerator as a producer. More specifically, the host program first prepares the input data for a computing task, which could involve copying the data across different memory spaces if necessary, before launching an execution kernel or an operation command in the accelerator. Then the accelerator continuously handles the tasks it is given, producing outcomes for the host program that keeps on querying for them. This program model has been supported by OpenCL [20] and CUDA [1]. It also allows us to abstract the accelerator as a separate computing unit when designing resource management in HETEE.

**PCIe based interconnect protocol**. Although new interconnect protocols to link computing units together have been proposed, such as NVIDIA NVLink, ARM CCIX, and IBM CAPI, PCIe is still the first or the only option for almost all commercial high-end heterogeneous computing systems (those involving GPUs, FPGAs, or TPUs), due to its extensive use for connecting devices to CPU. This is critical for heterogeneous computing, which is characterized by the intensive CPU-accelerator interactions. The design of HETEE takes full advantage of PCIe and uses a switch to dynamically allocate computing units across different secure enclaves.

**Software support for effective CPU-accelerator interactions**. Important to heterogeneous computing is effective CPU-accelerator interactions, which are supported by a software stack running on CPU. Taking GPUs as an example, its software stack includes GPU drivers, CUDA, and TensorFlow or Caffe. Another example is the domain-specific accelerator like compression, whose operations are enabled by driver, OpenCL etc. Therefore, this software stack brings in complicated communication between CPU and accelerator, which could leak out information once exposed to the untrusted OS, even when the communication is encrypted.

### 2.2 TEE Technologies

A trusted execution environment (TEE) guarantees the code and data loaded inside an isolated area (called an enclave) to be protected with respect to confidentiality, integrity and authenticity. TEEs aims to thwart sophisticated software adversaries (e.g. compromised OS) or even hardware adversaries who have physical access to the platform. TEE provides hardware-enforced security features including isolated execution, protecting the integrity and confidentiality of the enclave, along with the ability to authenticate the code running inside a trusted platform through attestation:

- *Isolation*. Data within the enclave cannot be read or modified by untrusted parties.
- *Integrity*. Runtime states should not have been tampered with.
- *Confidentiality*. Code, data and runtime states should not have been observable by unauthorized applications. This is achieved, e.g. by encrypting the code or data that reside in the memory.
- *Authentication*. The code under execution has been correctly instantiated on a trusted platform. To prevent simulation of hardware with a user-controlled software, a hardware root of trust burnt into the TEE chip during manufacturing can be used to perform remote attestation.

Trusted execution environment has been studied for decades. Varies of trusted hardware technologies were designed for specific application scenarios, supporting (some of) the above security features. ARM TrustZone is a collection of hardware modules that can be used to conceptually partition a system's resources between a secure world, which hosts a secure container, and a normal world, which runs an untrusted software stack. Intel's Software Guard Extensions (SGX) implements secure containers for applications without making any modifications to the processor's critical execution path.

## 2.3 PCIe Switch Fabric

PCIe ExpressFabric (switch) is an important interconnect technique for achieving cost-effectiveness in today's data center. This technique is leveraged in our design of HETEE, which to our best knowledge, has never been done before.

Unlike PCI that uses a shared bus architecture, PCIe has a point-to-point topology, using the switch chips to dynamically route packets of data between a PCIe host (e.g., *root complex* connected to the CPU and the memory subsystem) and multiple devices (e.g., GPUs, TPUs). This enables flexible sharing of IO among multiple devices, and also allows multiple hosts to reside on a single PCIe-based network using standard PCIe enumeration. These hosts communicate through the Ethernet-like DMA with each other and standard end-points, through unmodified applications. The PCIe switch can be hot-swapped and power cycled in any order, providing real plug and play for devices. Besides the capabilities of the IO sharing within multiple hosts and high performance host-to-host and host-to-IO DMA, the PCIe ExpressFabric also has two unique features important to our design:

**Software-defined fabric**. The switch is built on a hybrid hardware/software platform that offers high configurability and flexibility with regards to the number of hosts, end-points, and PCIe slots. Its critical pathways have direct hardware support, enabling the fabric to offer non-blocking, line speed performance with features such as IO sharing. The chip has a dedicated port for management, through which an external management CPU (mCPU) can initialize the switch, configure its routing tables, handle errors, Hot-Plug events, and others. In this way, all the hosts connected by the switch only see what the mCPU allows them to see.

**Flexible topologies**. The switch eliminates the topology restrictions of PCIe. Usually, PCI Express networks must be arranged in a hierarchical topology, with a single path from one point to another. ExpressFabric allows other topologies such as mesh, IO Expansion Box with Multiple hosts, and many others.

## 2.4 Threat Model

**(Privileged) Software adversary**. HETEE defends against the adversary with full control of the software stack in the insecure world, including unprivileged software running on the host, the host OS, virtual machine monitor, the initial boot code and system management mode. Such an adversary can also mount a side channel attack e.g. analyzing network traffic. The design of HETEE greatly reduces the surfaces of the attack, compared with SGX, exposing nothing but the total computing time of a task or the interval between encrypted data uploads (from the hard drive to the secure world). However, a thorough analysis of the leaks is left for the future research. We assume that the system software of the *security controller* is trusted.

**Hardware adversary**. HETEE defends against the adversary who has physical access to the platform. Without dedicated tamper-resistant mechanism, the hardware adversary might mount code boot attacks or snooping attacks on the host memory bus and the PCIe bus. Thus, we assume the hardware implementation of HETEE to be tamper-resistant, which can be achieved using the techniques discussed in Section 3.7. Besides, we do not consider electromagnetic or power analysis, and fault attacks.

We also assume that accelerators are directly controlled by their firmware, as happened in today's mainstream GPUs, like those produced by NVIDIA [39]. In this case, malicious code running on an accelerator cannot stop a reset operation commanded through a trusted driver and the firmware, which will return the accelerator to a trusted state (Section 3). Finally, we consider that the under the fundamental reset or cold reboot, the firmware on an accelerator (such as GPU) can ensure that all micro-architecture statuses, including those of caches and memory, are cleared.

**Others**. We use standard encryption algorithms. As such any cryptanalysis attempts are out of scope. So are denial-of-Service attacks, since they are trivial to launch for the adversary who controls the entire host platform.

## 3 HETEE Design

In this section, we elaborate the design and security analysis of HETEE.

### 3.1 Overview

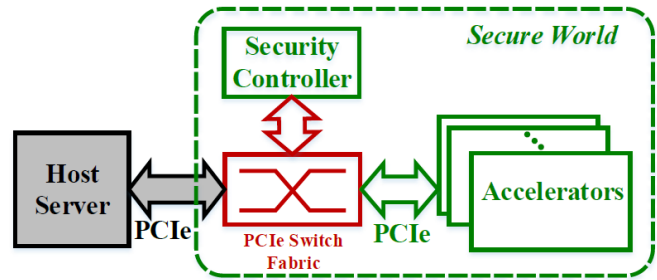

Figure 1: HETEE architecture.

**Design goals and principles**. HETEE is a computation-oriented TEE designed to achieve the following goals:

- *Strong isolation*. We expect that HETEE provides strong isolation both between secure/insecure worlds and between the enclaves that run different computing tasks. The isolation should be enforced not only logically but also physically to minimize the interface of the protected computing task exposed to the unauthorized party (e.g., a different computing task).

- *Flexible resource sharing*. We expect that different computing units (accelerators) can be dynamically allocated to

a computing task under protection (running inside an enclave). They should also be moved between the secure/insecure worlds during runtime without undermining the protection of sensitive data and computing tasks.
- *Thin TCB*. The TCB of HETEE system should be as thin as possible. Computing-unrelated components should be removed from its software stack. On the hardware side, only the *security controller* should be trusted.
- *Sytem-level implementation*. HETEE should be implemented on the system level, built upon *existing* accelerators, without changes to the commodity chips, which is critical for the technique's easy deployment.

**Architecture**. Here we present the *first* design for achieving the aforementioned goals, as illustrated in Figure 1. At the center of the design is a trusted *security controller* unit that manages a set of PCIe switches attached to a host server. The server runs on standard, commercial hardware with a standard OS, and the user-land programs supporting HETEE. *Controller* operates outside the untrusted OS, on top of its own software stack custom to CDI computing protection. Also connected to its PCIe switches are a set of computing units – accelerators.

*Security controller* and the accelerators (which are commercial, unmodified computing units) are encapsulated into a tamper-resistant box, which is connected to the host server through the PCIe switches. However, logically, only *controller* is considered trusted. Itself and the computing units under its protection form the *secure world*, and all other components (the host and unprotected units) constitutes the *insecure world*. Note that the accelerators are shared across these two worlds and only trusted when they are sanitized (through reset) and under the control of *security controller*.

The key idea of our design runs *security controller* as a stand-alone computing system to dynamically adjust the boundary of between secure and insecure worlds through the PCIe switches, rendering the control of an accelerator to the host OS when it is not needed for secure computing, and shifting it back when it is. To this end, we need to trust the custom OS and accelerator runtimes, drivers, etc. operating in *controller*, together with its encryption, authentication and remote attestation components. The host server and other computing systems communicate with *controller* through an in-memory task queue that accommodate the computing tasks offloaded to HETEE, in the form of task containing encrypted and signed code/binary or data. Also, we offer a generic and efficient programming interface for the host system, and a secure encrypted communication mechanism.

A remote user uses the remote attestation to verify the security of an accelerator. The main part of an offloading application runs on the CPU. It accepts the protection of a CPU TEE (e.g. Intel SGX), or its sensitive code and data exists in encrypted form. When the application needs to use an accelerator, it writes the code and data according to the agreed programming model, and sends these key data and task instructions to the *security controller* in ciphertext. These ciphertext are decrypted and parsed, and the tasks will be scheduled by a predetermined policy. The *security controller* prepares task-relevant data to the accelerator, and runs specified task commands to direct it to work until the result is retrieved, encrypted, and transmitted to the CPU application. Taking the NVIDIA GPU as an example, a CPU offloading application encrypts and signs the CUDA code and data and sends it to the task queue. The *security controller* decrypts and verifies these ciphertext on the trusted OS. Then, it runs CUDA codes locally and pass the data to the GPU for processing. Finally, the processing result is encrypted and copied to the target area of the CPU memory space specified by the task, and the CPU application is notified for further processing.

## 3.2 Security Controller

As a separate system, *security controller* runs on a board connected to the standard PCI fabric, acting as a gatekeeper for a secure world dedicated to CDI computing. It is the **only** unit that needs to be trusted. It runs a thin software stack to operate accelerator drivers and computing platforms like CUDA, together with secure computing functions like encryption/decryption, authentication and attestation (to the data holder).

**Software**. Linux is customized as trusted OS running on *security controller*, and the task level isolation is based on the existing container mechanism. Application-layer software mainly includes the following functions:

- *Task manager:* It is in charge of the task queue management, execution environment initialization, task parsing and dispatching etc.
- *Security task encryption/decryption, authentication and remote attestation:* the *controller* can establish an encrypted secure channel with the data holder through the intrusted host OS and collect measurement of a computing task to be performed on sensitive data in an enclave.
- *Accelerator runtime and driver.* These components are unmodified and operate just like in a normal OS.
- *Accelerator security mode switch. Security controller* is also the mCPU for PCIe ExpressFabirc chip of HETEE, which can dynamically configure PCIe fabric switching accelerators between secure and insecure worlds.

**Hardware**. The *security controller* is a stand-alone computing system, with its separate CPU, memory and disk[2]. It also has necessary hardware supports for secure boot. Here are the requirements for the hardware:

- The *security controller* should have good single-threaded performance and throughput capability.

---
[2]Here the disk could be shared with the untrusted OS in the presence of an interface built in the OS of the secure world.

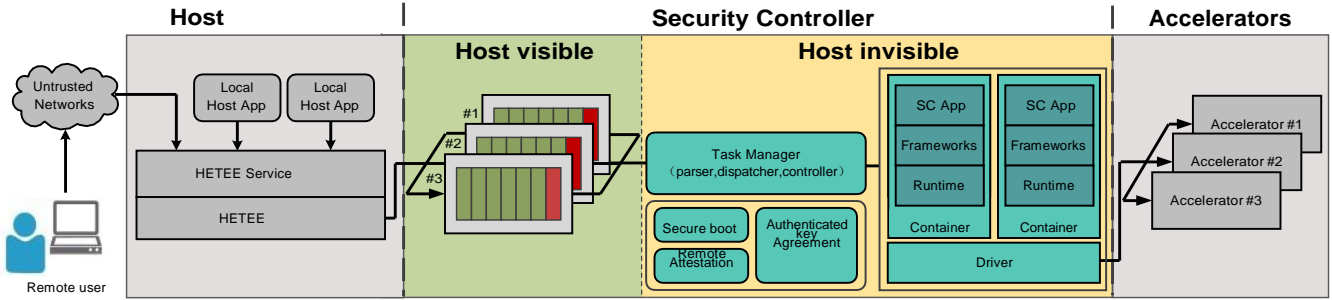

Figure 2: Task queue and software stacks on host and *security controller*.

- Its CPU should have strong IO bandwidth and processing capacity;
- Hardware acceleration engine for encryption and decryption and hash algorithms are expected.

In our research, we built the *security controller* on Xilinx ZC706 FPGA as a proof-of-concept for evaluating our designs for different HETEE components. A full-fledged implementation can customize PCB board with X86 processor and (or) FPGA chip, so that provides more computing power and effective supports for managing different accelerators.

### 3.3 Task Queue and Secure Programming Interfaces

Compared with the classical heterogeneous computing system, HETEE System introduces one more address space for the *security controller*. Having multiple address spaces presents a challenge for the design of programming interface. In addition, HETEE needs to provide a concise, generic programming interface that accepts security computing requests and data from remote or local hosts, and unifies them into HETEE enclave. It is important to note that this programming interface is completely independent on the CPU's TEE. More precisely, HETEE constructs a unified enclave for all compute units to be protected, including CPUs and accelerators.

We propose a task-based programming interface for HETEE system. This model divides the entire life-cycle of using the accelerator into three components: configuration task, command task, and data task.

- Configuration task: Host program sends the configuration task to the *security controller* to request the assignment of a new task queue, including the type and number of accelerators. Task manager assigns a uniform task ID to each procedure of using an accelerator.
- Command task: This type of task is primarily used to transfer programs to be executed on the *security controller*, which may be CUDA codes, or OpenCL programs, or other program code based on the accelerator API. To ensure the confidentiality and integrity, these codes are encrypted and signed before being encapsulated as a command task.
- Data task: This type of task is used to transmit sensitive data. These data may be encrypted inside the CPU TEE, or placed directly on the host hard disk in the encrypted form, or come from a remote user.

As shown in Figure 2, a set of task queues are provided between the *security controller* and the host CPU to support the task-based programming interface. The task queue is a memory data structure that belongs to the *security controller* space and is mapped into the host memory space by the standard PCIe protocol. We assign a task queue for each procedure of using the accelerator, in which all commands and data tasks have the same task ID and are fully sequenced (order are kept). These tasks are sent by the local or remote users, stored in the task queue, and processed by the task manager in the *security controller*.

```
#Check if remote attestation between remote user and
#HETEE is finished
res = CheckIfRemoteAttestationSucceed();
IF res is True:
  commandTask = ReceiveCommandTaskFromRemoteUser();
  WHILE True:
    #Insecurity Part
    computeOnUntrustedAccelerator(localNonPrivateData);
    #Security Part
    SendCommandTaskToHETEE(commandTask);
    #Encrypted private data is already stored on Host
    dataTask = Package(localEncryptedPrivateData);
    SendDataTaskToHETEE(dataTask);
    results = RecieveDataTaskResults();
  END WHILE
END IF
```

(a) Example pseudo code running on the host.

```
INPUT: InBuffer, OutBuffer, Accelerators

InitDevice(Accelerators);
AllocateDeviceBuffer(Accelerators);
WHILE True:
  #Call Get API to get data from InBuffer
  data = Get(InBuffer,size);
  IF data is not empty:
    CopyDataToDevice(data, Accelerators);
    #Normal way to use Accelerators
    ComputeOnDevice(Accelerators);
    results = CopyDataFromDevice(Accelerators);
    #Call Put API to put computing results to OutBuffer
    Put(OutBuffer,results);
  END IF
END WHILE
```

(b) Example pseudo code which is transmitted in term of command task and run on the *security controller* to use secure accelerators.

Figure 3: Example pseudo code to use HETEE

Figure 3(a) describes an example of a typical host program using the task model to securely use an accelerator. If a trusted remote user wants to run an application on HETEE, he should send a configuration task to the *security controller*. The task manager then allocates a new task queue to the user and notifies the user when the task queue is ready. Once the establishment of the task queue is completed, the remote user encrypts the command (i.e., AI model), packs the ciphertext into a command task, and put it into the task queue. The task manager fetches ciphertexts from the task queue, verifies the integrity of the messages, then decrypts and parses them. If the parsed task is a command, task manager will initiate a corresponding process (i.e., Tensorflow program) to execute the command. At the same time, input and output buffers are created for the process. When the task manager detects a new data task, it dispatches the task to the corresponding process input buffers. After its data is processed by accelerator, the process places the computing result in the output buffer. Task manager is also responsible for scanning the output buffer, encrypting the contents, send them back to the host memory through the DMA engine in term of data task. Then host server returns the encrypted results to the remote user. When the application finished, the remote user sends back a notification to acknowledge the *security controller*. Finally, the task manager destroys the task queues and terminates the process.

A command task contains the source codes or the corresponding binary executable file. HETEE provides a set of debugging and development environments similar to traditional heterogeneous computing for users to write this part of the code, and then encapsulate the source codes or the executable file into a command task as needed. It is important to note that these codes are built on the runtime and drivers on real accelerators, and should be executed on the *security controller*. Figure 3(b) describes the pseudo source codes corresponding to a typical command task. First, the process initializes accelerators and allocates device buffers for data. Then, the process loads new data from the input buffer, and moving them into the memory space of the accelerator. Then, the computing kernel is launched. After the execution, the result is copied from the accelerators to the output buffer. The above operations are iterated until the end of a task.

### 3.4 Computing Isolation and Elasticity

Shown in Figure 4, HETEE provides strong isolation between secure/insecure worlds and the enclaves that run different computing tasks. It achieves isolation mainly through the following techniques:

- The network composed by PCIe ExpressFabric Chip is used as the centric interconnection hub of the system. The *Security controller*, the host, and the accelerators are all connected via the PCIe ExpressFabric, which enables the physical isolation. The ExpressFabric chip can support direct communication between the host and the *security controller*, and can also allocate shared accelerator resources between secure/insecure world and across different enclaves in the secure world. Thus the *controller* just acts as a gatekeeper, and a proxy for the host system to access the accelerators in the secure world.

- The only interface between the *controller* and the host is the task queue, which communicates in the format and protocol described in Section 3.3. This simple mechanism minimizes the interface of the protected computing task exposed to the unauthorized host. The host and the *security controller* essentially belong to the two separate systems connected to the ExpressFabric chip. In order to enable the host to access the task queue belonging to the *controller* space, it is necessary to map the addresses of task queues into the space of the host. ExpressFabirc is responsible for routing and translating the address across different memory spaces via recording the mapping relations. The two sides can then use the high-performance Host-to-host DMA engine embedded in the ExpressFabric chip for high bandwidth and low latency data transfers.

- The *security controller* provides a container-based protection mechanism within its trusted OS to isolate different enclaves that run different computing tasks. It further enhances the isolation effect within the trusted OS.

A key design of the HETEE system is to take full advantage of the functionality of the software-defined network provided by the PCIe ExpressFabric chip. In other words, *security controller* dynamically allocates shared IO devices between the host and the *controller* through the mCPU management port of the ExpressFabric chip, defining and managing the boundaries of the secure world based on the predefined policy. This achieves the elasticity in computing in the following two perspectives.

**(1) Elastic accelerator allocation in secure world**

Multiple accelerators form an accelerator resource pool, which can make a good trade-off between cost and utilization [13]. HETEE only allows multiple tasks to timing-share accelerators, rather than multiple tasks concurrently running on them. Coupled with our context cleaning mechanism, such a design can minimize side-channel leakages across tasks. At the same time, HETEE permits a task to dynamically use multiple accelerators for better performance based on task requirements and load conditions. When a user needs to utilize the accelerator resource, a request is sent to the *security controller*, which dynamically allocates a certain amount of accelerator resources to the user according to the usage in the resource pool and the user's demand. Data and control commands from the user are assigned by the *security controller* to a set of accelerators, and the user sees only one virtual accelerator. The number of accelerators assigned to user can be dynamically adjusted by the *security controller*.

**(2) Elastic accelerator mode switch across secure/insecure world**

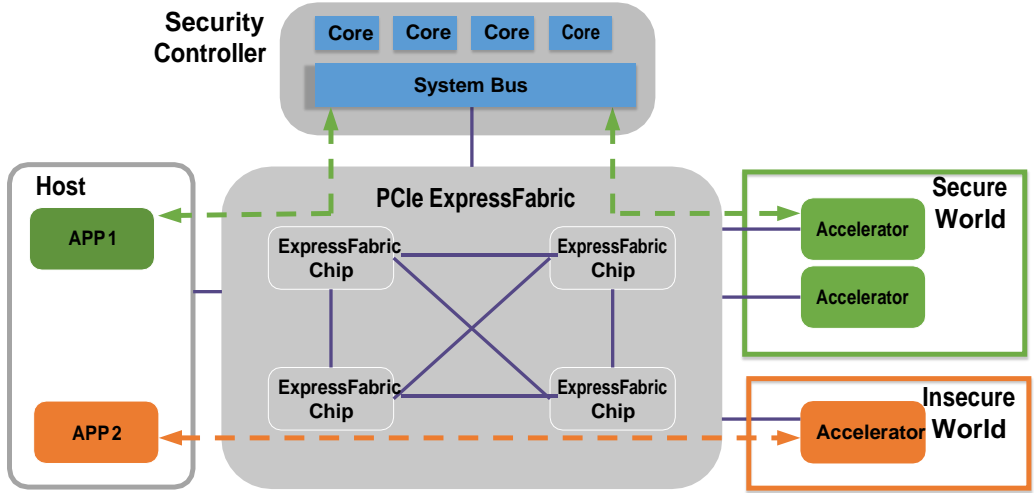

Figure 4: Accelerator resources isolation and elastic switch across secure/insecure world

In order to utilize the data center accelerator resources more efficiently, it is necessary to dynamically allocate the accelerators to different users, which might involve switching the accelerator across secure/insecure worlds. By dynamically configuring the ExpressFabric chip, an accelerator can be physically assigned to the host or the *security controller* on-demand, thus enabling the switching of the accelerator security mode. With this feature, HETEE dynamically defines the boundary of the secure world. To avoid the security risks from the data and code on the accelerator switched between different worlds, HETEE utilizes context clearance to restore the device to a trusted state. The detail is described in Section 3.5.

To support the security mode switching of accelerators, we have designed a preemptive accelerator scheduling mechanism based on priority. A secure switching service needs to be run on the *security controller*. In the meantime, the host can request the *controller* to release some of the accelerator resources from the secure world through the configuration task or an out-of-band request. The security switching service can also send high-priority requests to host, forcing a halt to the tasks performed by the accelerator in the insecure world, and take them dynamically back to the secure world.

## 3.5 Context Cleanup on Mode Switch

Prior studies have found that incomplete cleanup during a context switch exposes a large attack surface to adversaries [26,27,29,34,36,39–41,59]. Taking GPU chips as examples, Maurice et al. [36] demonstrated five different cleanup methods in virtualized environment, four of which still leaked sensitive data in the global memory. Therefore, a full and efficient context cleanup needs to be deployed to ensure the security protection of HETEE.

The complete *context* of an application includes not only the architectural, but also the micro-architectural states. Since implementations vary from vendor to vendor, the context includes, but is not limited to, the following contents.

Table 1: The overhead of different context cleanup methods on NVIDIA TITAN X GPU. (LoM:Local memory, ShM: shared memory, GlM:global memory)

| Context Cleanup Method | | Time Cost | Clear Size |
|---|---|---|---|
| API | *cudaDeviceReset* | 71ms | Resources of current process |
| | *cuCtxDestroy* | 53ms | Resources of current context |
| Software Reboot | nvidia-smi -r | 975ms | LoM/ShM/GlM, others are unknown |
| Code | Registers | 0.019ms | 24 × 65536 × 4B |
| | LoM | 50ms | 24 × 1024 × 512KB |
| | ShM | 0.020ms | 24 × 96KB |
| | GlM | 44ms | 12GB |
| | L1 Cache | 0.019ms | 24 × 48KB |
| | L2 Cache | 0.040ms | 3MB |
| Cold Reboot | Power off | ~minutes | All |

- *Architectural states:* the register files, on-chip scratchpad and memory, as well as off-chip memories.
- *Micro-architectural states:* on-chip caches, queues, and buffers etc.

However, it is noted that existing cleanup and reset mechanisms for commercial accelerators or GPUs usually cannot eliminate the risk of information leakage. In pursuit of performance, accelerator vendors usually provide quick reset and cleanup methods, which leave behind confidential context that pose a significant security risk [34]. In addition, some software invisible micro-architectures may be used to accelerate the processes [34,39,41], which indicates that software methods cannot clean up these resources.

As a case study, we investigate 4 cleanup approaches in-cluding software reboot, cold reboot, reset API and manually coded programs on latest NVIDIA GPU platforms with TITAN X and Maxwell architecture. Table 1 shows that cold reboot is the most effective but time-consuming method. A compromise is to combine the manual cleanup methods with API functions, which clean most of known context contents within acceptable time. A more thorough context cleanup requires the support of the hardware vendors. Fortunately, starting from Maxwell GPUs, NVIDIA introduces new mechanisms to strengthen security, and our experiments show TITAN X has good security features to be used during context switching.

Besides GPU, many accelerators are domain-specific and are designed into ASIC chips. Despite the wide variety of designs, the current designs of high-performance accelerators generally use multiple instances and shared On-chip caches. A thorough context cleanup needs to be fully considered.

### 3.6 Trust Establishment and Key Agreement

**Secure boot**. A secure boot scheme can be integrated to *security controller* to guarantee that it is running as expected. The firmware checks the signature of the boot software, including firmware drivers and the system software, and gives control to the system software only if the signatures are valid.

**Remote attestation**. A pivotal component of TEE is attesting to a remote data owner that a trusted software has been correctly instantiated on a trusted platform. We assume that a root secret (i.e. the private signing key) is burned into the e-fuses of the chip during the manufacturing process. We do not restrict on the choices of signing algorithms, however a Direct Anonymous Attestation (DAA) [5] protocol (e.g. EPID [6] used in Intel SGX) is necessary such that a device could prove to an external party without exposing the device identity.

The attestation process begins with the platform establishing a communication channel with the data owner who wishes to provision secret. Then the data owner sends an attestation request in the form of a randomly generated nonce. In the design of HETEE, only the measurement of the firmware as well as the system software is generated, since the state of accelerators is cleaned up upon a task switch. The measurement and the nonce along with the quote signed with the private signing key are sent back to the data owner. The data owner can verify the correctness of the signature and measurement.

**Authenticated key agreement**. A key agreement protocol (typically the Diffie-Hellman Key Exhange (DKE) [37]) can be integrated with remote attestation. It can be assured that a shared secret is established with a specific piece of software running on a trusted hardware. The shared secret will be used to encrypt the secret data for the data owner.

### 3.7 Security Analysis

In this section we present preliminary analysis of the security properties provided by HETEE.

**Protection against privileged software adversaries.** When configured as secure devices, the accelerators are registered to the address space of the *security controller* and are invisible to the host system. As such, HETEE enables physical isolation between the secure and insecure world, so that the privileged software adversary cannot directly access or tamper with the accelerators.

**Protection against physical adversaries.** HETEE environments are deployed in data center, but their administrators and operators are not trusted in our threat model. Even though the *security controller*, PCIe switches and accelerators are inside the same server blade, they are still subject to possible physical attacks. For instance, attacker might (in theory) observe the plain-text communication between *security controller* and accelerators in secure world, by using PCIe Logic analyzer to collect signals inside PCB. Thus it is necessary to build HETEE system with physical tamper resistance. Several mature techniques can be introduced in the design of HETEE.

- *Rack/Server/Adapter Enclosure*. Data centers often use flexible locking systems to manage access rights through flexible implementation methods, providing high reliability and high security. Combined with RFID readers [16], digital keypads [15], or fingerprint identification devices [14], access rights can be identified through a standalone locking system or a centralized network management system [47].
- *Protection Layer*. A protection layer is added to original circuits to prevent attackers from inspecting directly and reverse engineering. There are a variety of materials that can be used to implement the protective layer of the circuit, such as wire mesh [8], and doped circuits [4].
- *Self-destruction*. There are various research works proposing self-destruction methods to protect circuits from unwanted or unauthorized accesses. Tamper protection mesh [54] can be employed to recognize theses malicious accesses, such as opening the chassis or destroying protection layer [28]. Once detecting inspection, existing power supply [44] or offline power circuit [35] can be used to provide large current to destroy fusible circuits. Furthermore, electrically-conductive, self-destruct metal oxide films can burn the whole circuits when they are ignited [46].

**Reducing attack surfaces against the *security controller***. The *security controller* and the software running on it consists of the trusted computing base. HETEE is designed to reduce the attack surface of the *security controller*:

- *Controlled programming interfaces*. The programming interface is defined as task description and task queues. The format of tasks is predefined to avoid software attacks such as buffer overflow attacks. The tasks are parsed and processed on the accelerators and are unlikely to lead to privilege escalation on the *security controller*.

- *Cutting down the code base.* The software of the *security controller* is customized. It only needs to support limited hardware accelerators.

**Controlling side channels**. In the design of HETEE the states of accelerators are cleared upon mode switches and task switches. This reduces shared resources between the insecure world and secure world, as well as among tasks.

## 4 Implementation and Evaluation

### 4.1 Prototype System and AI Workloads

#### 4.1.1 Prototype System

Table 2: HETEE prototype system.

| GPU server with HETEE | Configuration |
|---|---|
| Host | SuperMicro GPU Server mother board (SYS-4029GP-TRT2), dual-socket Intel Xeon E5-2630v4 CPU, 512GB memory, 16TB Disk |
| PCIe fabric | PCIe Switch chips for IO expansion (transparent bridge) |
| Security Controller | FPGA system (Xilinx ZC706, which contains on-chip 600MHz ARM Cortex A9 dual cores, and 218.6K LUT, 437.2K FF programmable logics @100MHz) |
| Heterogeneous units | 4 GPUs (NVIDIA Tesla M40) |

The hardware platform includes a host CPU, a PCIe fabric, a *security controller* and several heterogeneous units. The detailed configuration is shown in Table 2. The Host CPU connects to the *security controller* and four GPUs via a PCIe Switch chip. The *security controller*'s implementation is the significance of our prototype system. The *security controller* includes two major functionalities.

*Security verification, task encryption and decryption, task analysis and queue management.* This part of the function mainly runs on the ZC706 FPGA chip. The ZC706 embeds an ARM Cortex A9 dual-core processor, where we run Linux as the secure OS of the *security controller*. All task requests are decrypted, parsed and dispatched within the Secure OS. In order to improve the efficiency of security authentication and en-decryption, we implemented the AES-GCM en-decryption coprocessor in the programmable logic part (PL side) of the ZC706. In addition, an efficient DMA engine is integrated for data transfer between host and ZC706 (Task communications between the Host CPU and the *security controller*).

*Tensorflow, CUDA library and GPU driver.* In the HETEE design, software stack of accelerator needs to be run in the trusted OS on *security controller*. In our prototype system, the ARM Cortex A9 processor of the ZC706 runs a Linux kernel. We did not find corresponding version of GPU driver and CUDA runtime for this configuration. Therefore, the prototype system did some tricky design. After the ZC706 decrypts and verifies the task sent from the host CPU, the data is transmitted back, and the GPU software stack including Tensorflow, CUDA library and GPU driver running on the host CPU are responsible for using the GPU. This implementation is appropriate since our prototype system is used for performance evaluation. Specifically, this implementation is more conservative than real HETEE in performance because of the extra FPGA-to-CPU communications.

In addition, the prototype system uses a PCIe transparent bridge as the interconnection center to connect the CPU, GPU and FPGA. The PCIe transparent bridge is a subset of the ExpressFabric chip. Because the embedded ARM A9 processor of the FPGA can act as both a slave and a master, the prototype system does not require the newest PCIe Express Fabric chip for performance evaluation experiments. In terms of performance, the PCIe transparent bridge's latency is basically the same as the PCIe ExpressFabric chip, thus the prototype system can meet the performance evaluation requirements. We are designing a HETEE hardware system based on a PCIe ExpressFabric chip, which is currently in the PCB production stage.

#### 4.1.2 AI workloads

We selected 6 classic neural networks, all of which almost obtained the best classification results at that time on the large-scale image classification dataset ImageNet [21, 30, 45, 48] and have been widely used in scientific research and industries. The number of layers in these networks are ranged between 8 and 152, with the number of parameters between 5 Million and 138 Million. Our choice is a good cover for typical neural networks from small scale to large scale. At the same time, we selected three different batch sizes as network inputs for each deep learning network to observe the impacts of input data of different sizes on the throughtput and latency of HETEE system. Table 3 describes the details of each network model.

### 4.2 Performance Evaluation

Compared to normal heterogeneous calculations, HETEE system introduces longer task (data and command) transfer paths, additional encryption and decryption and task processing work. This section mainly evaluates the throughput of the HETEE hardware prototype system running a set of deep learning network, and the latency of single-batch image reasoning (inference), as well as its scalability.

Table 3: AI workloads.

| Model | Parameters | Layers | Size of each image | Classes |
|---|---|---|---|---|
| AlexNet [30] | 60 Million | 8 | 227x227x3 | 1K |
| VGG16 [45] | 138 Million | 16 | 224x224x3 | 1K |
| GoogLeNet [48] | 5 Million | 22 | 224x224x3 | 1K |
| ResNet50 [21] | 25 Million | 50 | 224x224x3 | 1K |
| ResNet101 [21] | 44 Million | 101 | 224x224x3 | 1K |
| ResNet152 [21] | 60 Million | 152 | 224x224x3 | 1K |

**Baseline**: Normal unprotected GPU server is selected as the baseline, which has the same configurations with HETEE system (same CPU, memory, chipsets and GPU).

**ModelName_Batchsize**: the naming rule for AI network loads is ModelName_Batchsize. For example, VGG16_b32 means this network load use VGG16 model and its batch size is 32.

#### 4.2.1 Throughput and latency evaluation on single GPU

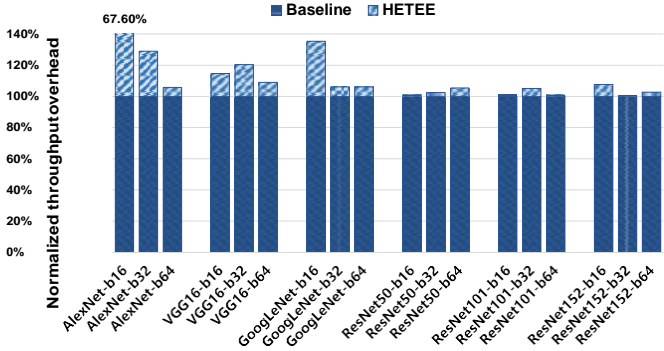

(a) Inference

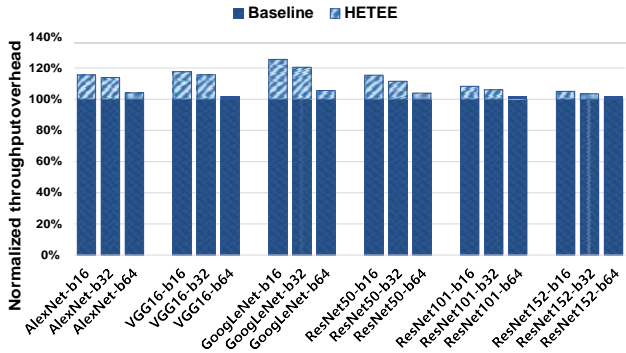

(b) Training

Figure 5: HETEE throughput evaluation on a single GPU.

The throughput evaluation of the HETEE system is shown in Figure 5 (the results of the HETEE are normalized to baseline). It can be observed that the HETEE throughput overhead is 12.34% for network inference and 9.87% for network training on average. As can be found from our evaluation, the batch size is an important factor that affects performance. In general, the appropriate increment of batch size can compensate for the overhead of task encapsulation, transmission, and processing. For most of network loads in training scenarios, their throughput can take benefits from larger batch size. This phenomenon is more stable in the training scenario. The reason is that computation in training is usually more intensive than inference. On the other hand, larger batch size means better bandwidth utilization. Balanced data transfer and processing usually lead to better throughput.

However, for other cases like VGG16 (inference), when batch size is increased from 16 to 32, bandwidth loss is positively correlated with data transfer time. When batch size is from 32 to 64, the GPU computing time becomes longer. At this time, the performance bottleneck is mainly the computing power of the GPU, the time of data transmission is partly obscured by GPU operation time, and thus the bandwidth loss is reduced.

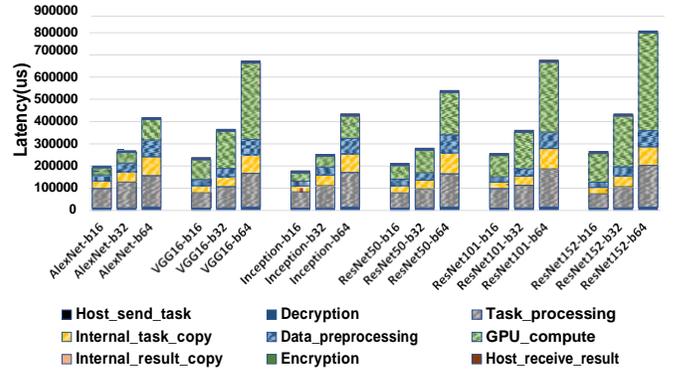

Figure 6: HETEE Latency of one inference task on a single GPU.

Figure 6 describes the latency and breakdown of the HETEE system for one inference of deep learning networks. As can be seen, the latency of the HETEE system increases by an average of 121.4% compared to the baseline system (baseline latency is the sum of GPU_compute and Data_preprocessing). Latency breakdown can help us with further analysis. First, a portion of the delay is unavoidable because the HETEE system introduces additional data transfer paths (including Host_send_task, Host_receive_result), as well as task parsing (part of Task_processing). Secondly, because of the limitations of our experimental platform, the ARM processor on the FPGA cannot run CUDA, and tasks need to be transferred between the FPGA and the CPU (including Internal_task_copy, Internal_result_copy). This part occupies the 14.6% of the total time on average. The real HETEE system can completely avoid this part of the cost. Finally, part of the time is spent on task transfer and wait between multiple queues (Task_processing). This part can be further optimized by introducing a faster interrupt mechanism.

#### 4.2.2 Scalability evaluation

The HETEE system supports the elastic allocation of accelerator resources. Multiple accelerators can be dynamically assigned to speedup the same enclave. Table 4 shows the scalability of HETEE. Two main conclusions can be drawn: (1) The elastic resource allocation mechanism of HETEE has good scalability. For most workloads, multiple GPUs achieve acceleration compared to a single GPU. Overall, the average acceleration ratio of 2GPUs relative to a single GPU is 1.51 on average, and 4GPUs acceleration is 2.16 on average; (2) HETEE does not affect its scalability compared to the baseline GPU Server. Taken ResNet152 as an example, the scalability of the HETEE system remains essentially the same as the baseline system. In addition, we also observed that for AlexNet-b16 and GoogLeNet-b16 network, the performance of 4GPUs is even worse than the performance of 2GPUs. This is mainly because of the GPU computing time is too short, and that the traffic between multiple GPUs and CPU is easy to cause GPU starving.

### 4.3 Cost Evaluation

In this section, we use several mainstream commercial GPUs as examples to analyze the cost of HETEE. We choose an Intel Xeon E5-2600 processor as the *security controller* considering its IO bandwidth and performance, with reference to the configuration of the commercial GPU server [50]. The cost of HETEE comes mainly from the newly added security controller system, as well as the PCIe switch chip(s). HETEE is primarily used to connect multiple accelerators and to form a relatively independent system with them. Therefore, when we make the cost assessment, we mainly pay attention to the ratio between the security related modules (*secure controller* and PCIe switch chip(s), etc.) and accelerator cost.

Figure 7 presents the result of our hardware cost analysis. We enumerate the cost ratio of HETEE of different quantities and types of GPUs. Because that there are 97 lanes in each PCIe Switch chip, one chip can only connect up to 5 GPUs (x16). When more GPUs are connected to PCIe fabric, the number of switch chips will increase accordingly. For example, integrating 8 GPUs needs two PCIe switch chips and a *security controller*. The corresponding PCB cost also needs to be doubled. In this 8-GPU configuration, the total cost of HETEE is $2779.

It can be seen the relative cost of HETEE decreases as the cost of GPU increases. Considering that a large number of GPU devices are usually deployed in a data center, it's obvious that the cost of HETEE is much lower than that of GPUs.

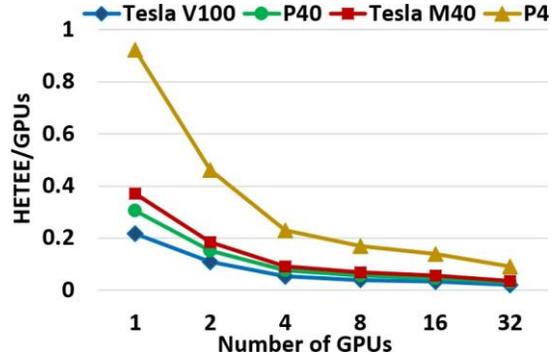

Figure 7: HETEE hardware cost analysis. (Security Controller: $999 [49], PCB: $300, Tesla V100: $8699 [51], Tesla P40 $4999 [51], Tesla M40: $5099 [51], Tesla P4: $2049 [51], PCIe Switch Chip: $590 [7])

## 5 Discussion

**From off-chip to on-chip *security controller*.** Existing HETEE design considers maximizing compatibility with existing CPU and accelerator chips, and CPU and *security controller* interconnect via off-chip PCIe Expressfabric. However, heterogeneous architectures are also widely used inside SOC chips, such as many application processors or high-end chips that integrate a variety of accelerator in the chip. One of the main differences between on-chip and off-chip heterogeneous architectures is the different interconnect buses. Heterogeneous SOC chips mainly use AXI or other bus protocols, it is clear that existing HETEE design needs to be changed. We consider using a multi-level bus, where the CPU and heterogeneous acceleration units are connected by a bridge (HUB) that can be dynamically configured. *Security controller* is also integrated into the chip as a standalone system and can control the hub that connects the CPU to the heterogeneous unit. With the exception of different interconnection hub options, most of the rest of the HETEE design does not need to be changed.

**Thin TCB of trusted Linux on *security controller*.** Building trusted OS on *security controller* requires a lot of engineering and research work to do. This part of the detail is beyond the scope of this article, and we will focus on it in the future. First, we need to customize the Linux kernel and delete all parts that are not related to task processing. Second, security modules and security review mechanisms need to be introduced inside trusted OS. Specific can be learned from LSM [56], and SELinux [31]. Finally, we use the container mechanism within the trusted OS to isolate different computing tasks. At the same time, *security controller* should efficiently run TensorFlow, CUDA, OpenCL, and a variety of drivers for kinds of accelerators. How to make a better trade-off between security and performance is one of the key tasks for the future.

Table 4: HETEE scalability throughput evaluation on multiple GPUs for inference tasks (For all cases with the same model and batch size, their throughputs are normalized to the case of baseline with 1 GPU)

| Model | Batch size | 16 | | | 32 | | | 64 | | |
|---|---|---|---|---|---|---|---|---|---|---|
| | Number of GPU | 1 GPU | 2 GPUs | 4 GPUs | 1 GPU | 2 GPUs | 4 GPUs | 1 GPU | 2 GPUs | 4 GPUs |
| AlexNet | Baseline | 1.00 | 1.24 | 1.57 | 1.00 | 1.30 | 1.74 | 1.00 | 1.68 | 2.20 |
| | HETEE | 0.60 | 0.62 | 0.61 | 0.77 | 1.16 | 1.28 | 0.94 | 1.62 | 1.74 |
| VGG16 | Baseline | 1.00 | 1.87 | 2.62 | 1.00 | 1.78 | 2.91 | 1.00 | 1.94 | 3.34 |
| | HETEE | 0.87 | 1.77 | 2.40 | 0.83 | 1.70 | 2.76 | 0.92 | 1.78 | 3.04 |
| GoogLeNet | Baseline | 1.00 | 1.35 | 1.61 | 1.00 | 1.41 | 1.87 | 1.00 | 1.35 | 1.89 |
| | HETEE | 0.74 | 0.82 | 0.80 | 0.94 | 1.34 | 1.46 | 0.94 | 1.26 | 1.65 |
| ResNet50 | Baseline | 1.00 | 1.68 | 2.72 | 1.00 | 1.56 | 2.57 | 1.00 | 1.70 | 2.63 |
| | HETEE | 0.99 | 1.67 | 1.65 | 0.98 | 1.51 | 2.30 | 0.95 | 1.58 | 2.45 |
| ResNet101 | Baseline | 1.00 | 1.85 | 3.23 | 1.00 | 1.74 | 2.91 | 1.00 | 1.72 | 2.93 |
| | HETEE | 0.99 | 1.81 | 2.56 | 0.95 | 1.62 | 2.71 | 0.99 | 1.65 | 2.70 |
| ResNet152 | Baseline | 1.00 | 1.84 | 3.31 | 1.00 | 1.85 | 3.28 | 1.00 | 1.81 | 3.17 |
| | HETEE | 0.93 | 1.76 | 2.90 | 0.99 | 1.77 | 2.97 | 0.97 | 1.73 | 2.97 |

## 6 Related Work

**Isolated Execution**. Mainstream processor vendors have implemented TEEs in some of their chip products, such as Intel Software Guard Extensions (SGX) [25], AMD Secure Encrypted Virtualization (SEV) [24] and ARM TrustZone [3]. In addition, based on the concept of "open-source security", Keystone [33] and Sanctum [12,32] are proposed to design an open-source secure enclave for RISC-V processor. These TEEs generally isolate a secure world from the insecure one, and the protected data can be processed in such secure world. However, none of those TEEs can truly support CDI computing tasks which widely adopt heterogeneous architecture. For example, Intel SGX does not support trusted IO paths to protect the data transmissions between enclaves and IO devices. Although ARM TrustZone can support trusted IO paths for some certain peripherals in ARM ecosystem, it is noted that TrustZone still does not truly support heterogeneous computing, especially for off-chip accelerators like GPU, FPGA and other ASICs etc.

**Trusted paths**. Graviton [53] is an architecture that supports trusted execution environments on GPU. It needs to modify the existing GPU chips via enhancing the internal hardware command processor. As a comparison, HETEE does not require any changes for existing commercial CPUs or accelerator. SGXIO [55] is a generic trusted path extension for Intel SGX. Trusted paths are established via a dedicated trusted boot enclave. However, the capacity limitation of SGX enclave prevent SGXIO being widely used for high-performance accelerators like off-chip GPU. Besides, there are several work which propose specific trusted paths for some certain of accelerators. For instance, Zhou et al. proposed a trusted path to a single USB device [58]. And Yu et al. demonstrated how to build trusted path for GPU separation [57]. Filyanov et al. discussed a pure uni-directional trusted path using the TPM and Intel TXT [17].

**Privacy preserving deep learning**. Nick Hynes et al evaluated two types secure AI computing scenes [23]. One scenario is to compute the entire AI workloads inside SGX enclaves. Obviously, this method can not achieve the efficiency of using specific accelerators. Another scenario they evaluated was the Slalom solution [52], which used trusted hardware in tandem with an untrusted GPU. Slalom needs to decompose the AI model network into two parts, in which the upper control flow part runs inside the SGX enclave and is tightly protected, while, like convolution, matrix multiplication of these non-privacy-sensitive basic computations is thrown to untrusted GPU to accelerate computation. However, splitting AI networks will result in a decrease of training and inference accuracy. While our HETEE programming model uses a set of special APIs to securly encapsulate the whole AI network, without change the internal structure of the deep learning model, so that the calculation accuracy will not be affected.

## 7 Conclusion

Privacy-preserving techniques capable of supporting compute- and data-intensive (CDI) computing are important in the era of big data. Emerging as a more practical solution is the new generation of hardware supports for TEEs. However, none of those existing TEEs can truly support CDI computing tasks, due to their exclusion of high-throughput accelerators like GPU and TPU. This paper present the HETEE (Heterogeneous TEE), the first design of TEE capable of strongly protecting heterogeneous computing. The proposed design enables collaborative computing units (CPU and heterogeneous units) to be protected under a single enclave and conveniently assigned across secure/insecure worlds and different enclaves. HETEE is uniquely constructed to work with today's servers, and does not require any changes for existing computing chips.

It includes a security controller running on a board connected to the standard PCI Express (or PCIe) fabric, acting as a gatekeeper for a secure world dedicated to CDI computing. We partially implemented HETEE on the prototype system which is a X86 host system connected with Xilinx Zynq FPGA (acting as the *security controller*) and NVIDIA GPUs over PCIe fabric, and evaluated it with large-scale Neural Networks inference tasks. Our evaluations show that HETEE can easily support such secure computing tasks with acceptable performance overhead and exhibits good scalability when elastically using multiple accelerators.